\documentclass[a4paper,12pt]{article}
\usepackage{amssymb}
\usepackage{epsfig}

\begin{document}
\title{Midplane Gas Density and the Schmidt Law}
\author{A.\,V. Zasov and O.\,V. Abramova
\thanks{Sternberg Astronomical Institute,
Universitetskii pr. 13, Moscow, 119991 Russia; {oxana@sai.msu.ru},
{zasov@sai.msu.ru}}}

\maketitle

\begin{abstract}
\hspace{0.6cm}The thickness of isothermal gaseous layers and their
midplane volume densities $\rho_{gas}(R)$ were calculated for
several spiral and $LSB$ galaxies by solving the self-consistent
equilibrium equations for gaseous discs embedded into a stellar one.
The self-gravity of the gas and influence of dark halo on the disk
thickness were taken into account. The resulting midplane volume
densities of spiral galaxies  were compared with the azimuthally
averaged star formation rate $SFR$ to verify the feasibility and
universality of the Schmidt law $SFR\sim \rho_{gas}^n$.
\end{abstract}
\hspace{0.6cm}Gas density is the major parameter which determines
the  star formation rate in galaxies. Scmidt~\cite{ZA_Schmidt59}
suggested a simple form of the ``volume'' star formation rate -- gas
density relationship: $SFR_v\sim\rho_{gas}^n$ (where $n\approx 2$
for the solar vicinity), usually called the Schmidt law. Being
essentially empirical, the Schmidt law and its modifications open a
possibility to calculate the evolution models of galaxies,
parameterizing the star formation history. However, the power index
$n$ cannot be found directly from observations of other galaxies,
because in order to estimate $\rho_{gas}$ it is necessary to know
the gas layer thickness, which may vary significantly both along the
galaxy radius and from one galaxy to another. Therefore in practice
the Schmidt law is often replaced by the other one, superficially
similar empirical law $SFR_s\sim\sigma_{gas}^N$ (usually called the
Kennicutt\,--\,Schmidt law), where the compared values are scaled to
unit disc surface area. In most cases values of $N$ obtained for
different galaxies lay within the limits of $1<N<2$  but for some
galaxies they prove to be much steeper ($N>3$ for M~33, see ~Heyer
et al.~\cite{ZA_HeyerAll04}).

To estimate $\rho_{gas}$, in this work we used the data on the gas
surface density distributions, brightness distributions and velocity
curves for galaxies taken from the literature. By solving the
equilibrium and Poisson equations for stellar, $HI$ and $H_2$ discs,
we calculated the midplane gas and star volume densities as a
function of the radial distance $R$ for several spiral and (for
comparison) $LSB$ galaxies. The former include: M33, M51, M81, M100,
M101, M106 and our Galaxy. The central parts of galaxies where the
bulge dominates and/or the observed rotation curve is uncertain were
ignored. We assumed that the stellar and gaseous discs are
axisymmetric  being in hydrostatic equilibrium and that the pressure
of gas is determined by its turbulent motion:
$P_{gas}=\rho_{gas}\,C_{gas}^2$, where the velocity dispersion
$C_{gas}$ was taken to be constant (although different for atomic
and molecular gas). Both the self gravitation of gas and presence of
dark halos which also influence the thickness of discs were taken
into account. The equations were numerically solved using an
iterative algorithm (see the details in Zasov \&
Abramova~\cite{ZA_AZ}).

To estimate the stellar disc thickness, two models  were employed:
stellar velocity dispersion $C_z$ was assumed to be either a
constant or proportional to the marginal radial velocity dispersion
$C_r$, that provides stellar disc with gravitational stability (see
the discussion in \cite{ZA_Bottema93,ZA_ZasAll04}). Both models give
rather similar results.

\begin{figure}[!ht]
\epsfig{file=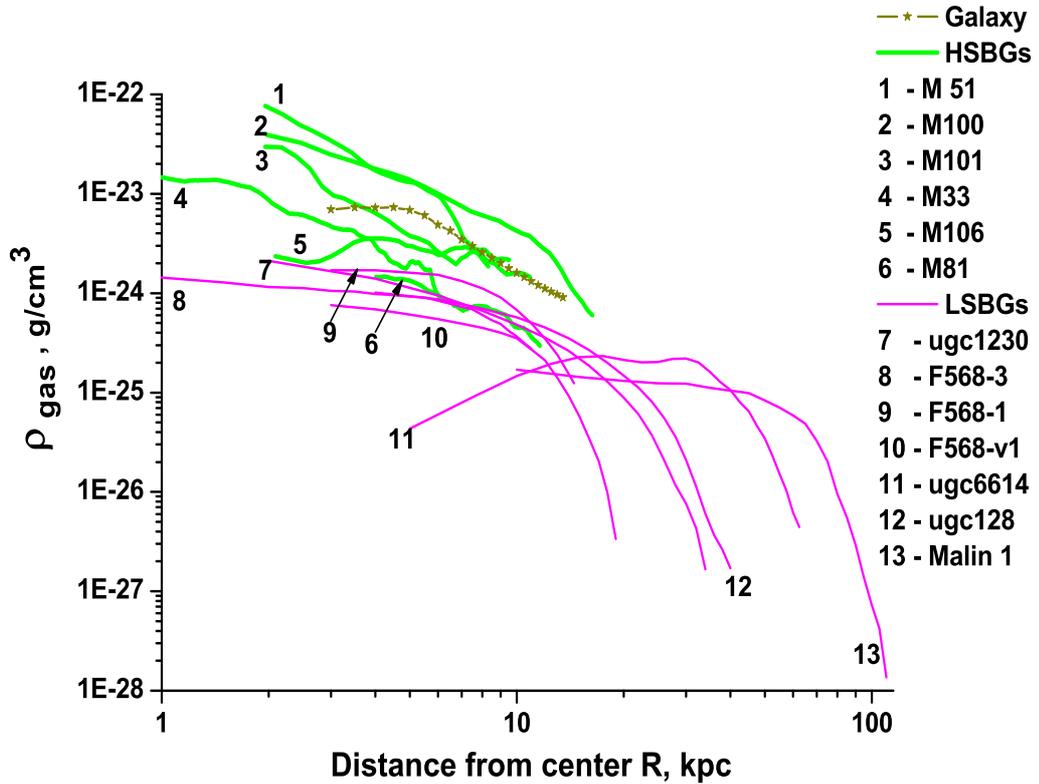,height=12cm,width=15cm} \caption{Midplane
gas density distributions in chosen galaxies.} \label{fig1}
\end{figure}

The obtained estimates of the midplane gas density $\rho_{gas}(R)$
for spiral and $LSB$ galaxies are illustrated in Fig.~\ref{fig1}.

To compare the surface $(SFR_s)$ and volume $(SFR_v)$ star formation
rates with the gas densities in spiral galaxies we used the
estimates of $SFR_s$ from Boissier et al.~\cite{ZA_main}, based on
the smoothed absorption-corrected UV profiles. The resulting
diagrams are demonstrated in Figs.~\ref{fig1}\,a,b.

\begin{figure}[!ht]
\epsfig{file=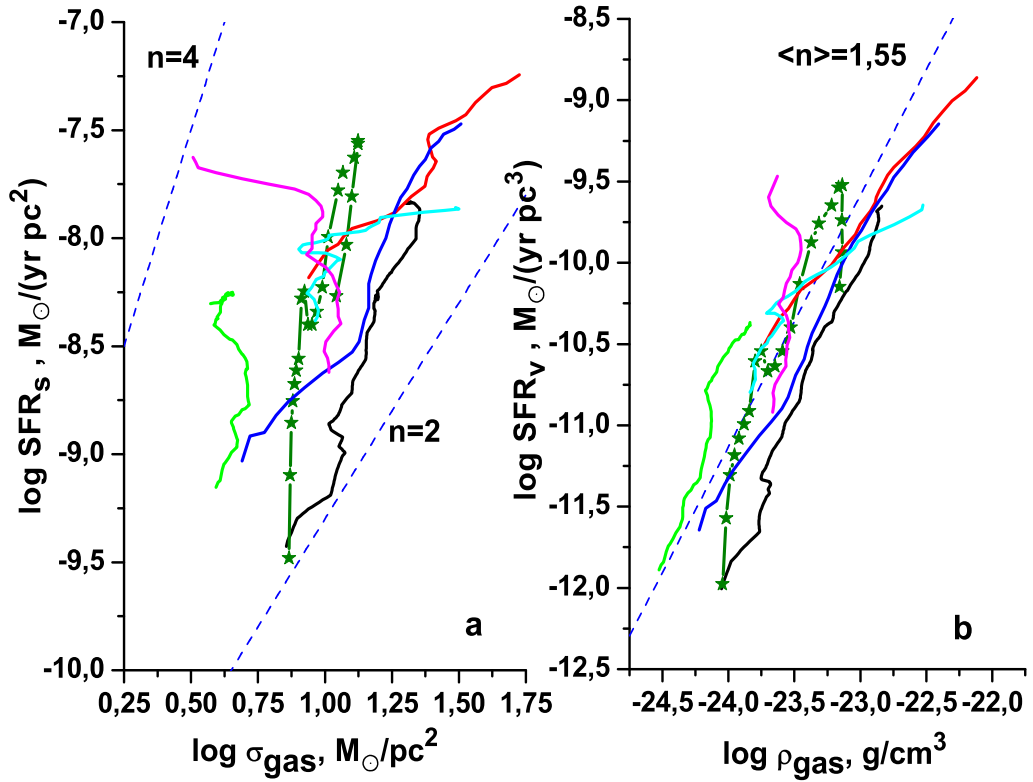,height=12cm,width=15cm} \caption{The Schmidt
laws: $SFR_s$ vs. the surface gas density $(a)$ and $SFR_v$ vs. the
volume gas density $(b)$ for spiral galaxies.} \label{fig2}
\end{figure}

{\bf Main results}.

\emph{\textbf{1}}. Marginally stable stellar discs in all cases
but M33 and our Galaxy increase their thickness significantly
beyond $R\approx 2-3\,R_0$. \emph{\textbf{2}}. Gaseous discs of
$LSB$ galaxies are thicker than the stellar ones, and $\rho_{gas}$
is about an order of magnitudes lower than in HSB spiral galaxies.
\emph{\textbf{3}}. There is no universal Schmidt law
$SFR\sim\rho_{gas}^n$, common to all galaxies. Nevertheless,
$SFR$, taken for the whole complex of galaxies, reveals better
correlation with the volume gas density than with the column one.
\emph{\textbf{4}}. Parameter $n$ in the Schmidt law in spiral
galaxies ranges between 0.8 (M101) and 2.4 (M81). However if to
consider the molecular gas only,  the mean value of $n$ becomes
close to unit.

\bigskip
 This work was supported by the Russian Fond of Basic Researches grant 07-02-00792.

\end{document}